\documentclass[conference]{IEEEtran}

\usepackage[pdftex]{graphicx}
\usepackage{amsmath}
\usepackage{comment}
\usepackage{eqparbox}
\usepackage{tabularx}
\usepackage{longtable}
\usepackage{cite}
\usepackage [english]{babel}
\usepackage [autostyle, english = american]{csquotes}
\MakeOuterQuote{"}

\begin{document}

\title{Leveraging Natural Language Processing to Mine Issues on Twitter During the COVID-19 Pandemic\\
}

\author{\IEEEauthorblockN{Ankita Agarwal\textsuperscript{\textsection}
\IEEEauthorblockA{\textit{Department of Computer Science} \\
\textit{Wright State University}\\
Dayton, Ohio 45435 USA \\
agarwal.15@wright.edu}
\and
\IEEEauthorblockN{Preetham Salehundam\textsuperscript{\textsection}}}
\IEEEauthorblockA{\textit{Department of Computer Science} \\
\textit{Wright State University}\\
Dayton, Ohio 45435 USA \\
salehundam.2@wright.edu}
\and
\IEEEauthorblockN{Swati Padhee}
\IEEEauthorblockA{\textit{Department of Computer Science} \\
\textit{Wright State University}\\
Dayton, Ohio 45435 USA \\
padhee.2@wright.edu}
\and
\IEEEauthorblockN{William L. Romine}
\IEEEauthorblockA{\textit{Department of Biological Sciences} \\
\textit{Wright State University}\\
Dayton, Ohio 45435 USA \\
william.romine@wright.edu}
\and
\IEEEauthorblockN{Tanvi Banerjee}
\IEEEauthorblockA{\textit{Department of Computer Science} \\
\textit{Wright State University}\\
Dayton, Ohio 45435 USA \\
tanvi.banerjee@wright.edu}
}

\maketitle

\begingroup\renewcommand\thefootnote{\textsection}
\footnotetext{co-first authors}
\endgroup


\begin{abstract}
The recent global outbreak of the coronavirus disease (COVID-19) has spread to all corners of the globe. The international travel ban, panic buying, and the need for self-quarantine are among the many other social challenges brought about in this new era. Twitter platforms have been used in various public health studies to identify public opinion about an event at the local and global scale. To understand the public concerns and responses to the pandemic, a system that can leverage machine learning techniques to filter out irrelevant tweets and identify the important topics of discussion on social media platforms like Twitter is needed. In this study, we constructed a system to identify the relevant tweets related to the COVID-19 pandemic throughout January 1\textsuperscript{st}, 2020 to April 30\textsuperscript{th}, 2020 and explored topic modeling to identify the most discussed topics and themes during this period in our data set. Additionally, we analyzed the temporal changes in the topics with respect to the events that occurred during this pandemic. We found out that eight topics were sufficient to identify the themes in our corpus. These topics depicted a temporal trend. The dominant topics vary over time and align with the events related to the COVID-19 pandemic.

\end{abstract}


\section{Introduction}

Since December 2019, a series of acute atypical respiratory disease outbreak in Wuhan, China, rapidly spread globally. It was soon discovered that a novel coronavirus, "SARS-CoV-2" was responsible for the outbreak. The disease caused by this virus was called Coronavirus disease 2019 (COVID-19), and a pandemic was declared by the World Health Organization (WHO). The disease was found to be highly contagious, with a reproduction rate of  2.2 (R0) \cite{Velavan2020-nn}. As of August 21\textsuperscript{st}, 2020, a total of 22,864,873 people were infected globally, and 797,787 deaths were confirmed in 188 countries \cite{dong2020interactive}.
With time, COVID-19 has turned out to be an economic and social crisis, attacking societal structure at its core. The crisis has affected all public segments, particularly detrimental to members of most vulnerable social groups like older adults and people with weak immune systems. According to the UN Department of Economic and Social Affairs (UN DESA\footnote{https://www.un.org/development/desa/dspd/everyone-included-covid-19.html}), if not adequately addressed the social crisis created by the COVID-19 pandemic may also increase inequality, exclusion, discrimination and global unemployment in the medium and long term. 

Social media are often seen as fast and effective platforms for searching, sharing, and distributing health information among the general population. They play a pivotal role in information dissemination and consumption during sudden outbreaks for people to access timely and reliable information about the disease symptoms and its prevention. 
However, social media has both positive and negative social impacts during the crisis, with increasing social distancing and growing reliance on online communication. Recent studies \cite{Jordan2018-vv,Shah2019-ik,Sinnenberg2017-nk,Shah2019-tj,Steffens2019-el} have shown that social media can play an essential role as a source of data for understanding public attitudes and behaviors during a crisis. Also, in the case of strong emotional reactions, media coverage of the pandemic may influence public sentiments\cite{van2020using}. 
Social media data can be used to quickly identify the main thoughts, attitudes, feelings, and topics that are occupying the minds of people about the COVID-19 pandemic. Such data can help policymakers, health care professionals, and the public identify primary issues that are of concern and address them in a more appropriate manner \cite{BELKAHLADRISS2019560}. Since January 2020, the number of papers analyzing Twitter activity during the COVID-19 pandemic has been increasing. Recent studies have identified the topics of discussions, concerns, and controversies surrounding COVID-19 in social media \cite{Liu2020-vx,Abd-Alrazaq_undated-ln,Dong_undated-wr,Chen2020-pn}.

In this study, we seek to understand the extent to which social media conversations are a part of the discourse for different events unfolding during a pandemic. We investigated the effect of the events which occurred during the COVID-19 pandemic on major topics of discussion on Twitter over time. 
We addressed the following research questions in this study:
\begin{enumerate}
\item 
Research Q1: What is the feasibility of differentiating between relevant and non-relevant tweets with respect to the COVID-19 pandemic?
\item
Research Q2: What are the trending topics in public discussions on Twitter related to the COVID-19 pandemic?
\item 
Research Q3: What is the relationship between events related to COVID-19 and the trending topics on Twitter over time? 
\end{enumerate}


\section{Related Works}
Understanding the prevalent topics during COVID-19 involved two major tasks: (1) Filtering the relevant tweets providing situational and actionable information about the pandemic, and  
(2) Analyzing the latent topics and themes with respect to the events which might have influenced public opinion. We provide a literature survey of each of these steps.  

\subsection{Relevant Tweet Extraction}
The preliminary research on analyzing social media data related to the COVID-19 pandemic has increased almost daily. Some works have focused on analyzing human behavior and reactions to the spread of COVID-19 in the online world \cite{Li2020-ne,Rashid2020-oe,Schild2020-nv,Abd-Alrazaq_undated-ln}, or investigating the conspiracy theories and social activism \cite{Ferrara2020-ra,Shahsavari2020-yh}. 
Often studies ignore the need to preprocess the tweets to filter out the tweets that do not pertain to the topic under investigation. For example, in a drug-related Twitter study \cite{lamy2017increases}, one of the keywords used to capture relevant tweets was the word “spice”. However, tweets related to pumpkin spice latte were collected, which were not relevant to the topic under investigation. Relevant tweets provide both situational and actionable information to extract requirement-specific information, decision-making, and immediate assistance to the affected people \cite{Vieweg2010-ua}. Several studies have mainly focused on identifying relevant information on Twitter. While \cite{Aramaki2011-ub} have identified tweets relevant to influenza, \cite{Muppalla2017-am} have worked on filtering out noisy tweets by identifying relevant tweets for the Zika virus using text-based features. Wahbeh et al. \cite{Wahbeh_undated-tu} have utilized a qualitative approach to identify relevant tweets during COVID-19. Although recent studies have used deep learning methods for social media mining in natural disasters \cite{inproceedings}, the same for pandemics is limited. In our work, we experiment with a range of machine learning and deep learning language models to filter out the relevant tweets for COVID-19.

\subsection{Topic Modeling and Analysis}
Topic modeling has been applied in areas like health informatics \cite{Zalewski2017-ga}, social media networks \cite{Ostrowski2015-uh}, telecommunication, and digital payments \cite{Agarwal2020-xx} in order to organize and summarize large textual information and to uncover hidden thematic structures in a collection of documents \cite{Blei2003-qr}. Similarly, in the case of crisis and pandemic scenarios, topic modeling has played an important role in understanding the most prevalent themes discussed on social media platforms. Recent works include performing topic modeling on documents related to disease outbreaks like the Ebola virus epidemic \cite{Lazard2015-to} , Zika virus epidemic \cite{Miller2017-lb}, dengue epidemic \cite{Missier2016-al}, and the seasonal influenza \cite{Kagashe2017-ms}.
Some studies have explored the feasibility of topic modeling on documents extracted from other social media platforms such as Facebook \cite{Raamkumar_undated-cy}, Weibo \cite{Han2020-kw} as well as Google trends \cite{Bhattacharya2020-xv}.  
Additionally, Medford et al. \cite{Medford_undated-zz}  studied the change in topics throughout January 14\textsuperscript{th} to 28\textsuperscript{th}, 2020 using Twitter data related to COVID-19. In addition, Ordun et al. \cite{Ordun2020-qh} explored a time series based analysis, and Yin et al. \cite{Yin2020-sw} explored Dynamic Topic Modeling to study the trending topics in tweets for the pandemic.
However, to the best of our knowledge, no work has discussed the influence of the events related to COVID-19 on the trending topics over time being discussed by the public on social media. We attempt to understand how public discussions on social media about COVID-19 resonate with the events related to this pandemic.

\section{Data}
\label{sec:data_collection}
In this section, we describe our data collection, hand-annotated data and preprocessing steps involved in data cleaning.

\subsection{Data Collection}
We utilized the COVID-19 pandemic-specific keywords like “coronavirus”, “covid-19”, “sars-ncov” to extract tweets posted from January 1\textsuperscript{st} 2020 to April 30\textsuperscript{th} 2020. In early January (January 9), the World Health Organization (WHO) announced cases of coronavirus-related pneumonia in Mainland China\footnote{https://www.who.int/china/news/detail/09-01-2020-who-statement-regarding-cluster-of-pneumonia-cases-in-wuhan-china}. The WHO officially declared COVID-19 a global pandemic on March 11\footnote{https://www.who.int/dg/speeches/detail/who-director-general-s-opening-remarks-at-the-media-briefing-on-covid-19---11-march-2020}.  By late April (April 29), the National Institute of Health (NIH) announced that the medication, remdesivir, performed better than placebo in treating COVID-19 \footnote{https://www.nih.gov/news-events/news-releases/nih-clinical-trial-shows-remdesivir-accelerates-recovery-advanced-covid-19}. An open-source crawler library, getOldTweets\footnote{ https://github.com/Mottl/GetOldTweets3} and Twitter Streaming API were used to extract the tweets. The getOldTweets library was used to overcome the daily quota limitations of the Twitter API. Additionally, this study was limited to English language tweets alone, and a total of 957,923 tweets were collected and preprocessed as described in Section \ref{sec:data_preproc}. Consequently, 866,527 unique tweets were obtained and serve as our dataset.

\subsection{Data Annotation}
\label{sec:data_annotation}

When tweets were collected using keywords, often, there were instances of the tweets not related to COVID-19 but still containing the specified keywords. In Table \ref{tab:sample_tweets}, we report sample tweets that contain the keyword "coronavirus", but are not relevant to the domain of COVID-19 pandemic. We randomly sampled 1,500 unique tweets out of 866,527 unique tweets, and three annotators labeled them as "Relevant" or "Irrelevant". A tweet was labeled as "Relevant" if it contained information about the spread, cause, effect, opinion, sentiment, emotion with regards to COVID-19, otherwise, it was labeled as "Irrelevant". Out of 1,500 tweets, 1,154 (77\%) tweets were labeled as "Relevant" and 346 (23\%) as "Irrelevant," which served as our dataset for building a relevancy classifier. 

\begin{table}[htbp]
\centering
\smallskip
\caption{Example of Relevant and Irrelevant Tweets}
\scalebox{1.3}{
\begin{tabular}{|p{4.15cm} |p{1.15cm}|}
\hline
\textbf{Tweet} &\textbf{Label} \\
\hline

another foolproof way of not dying of coronavirus is to stick your head in an oven lit on gas mark 8. I expect there are loads more such as being hit by an asteroid. &Irrelevant \\
\hline
coronavirus live updates: cases up nearly 60\% as airports expand screenings: ny times - &Relevant\\
\hline
\end{tabular}%
}
\label{tab:sample_tweets}
\end{table}
\subsection{Data Preprocessing}
\label{sec:data_preproc}

The tweets fed to the relevancy classifier and the topic modeling were first pre-processed, as discussed in this section. Usually, the language used in social media posts is very informal. People use short words, self-curated abbreviations, URLs, and multiple punctuations, periods, or exclamations. To understand the context and semantics of the content, we first converted the tweets to lowercase and then removed the emoticons, stop words, punctuations, and URLs, including symbols like “\#” ,“@” as well as special characters. We also removed the keywords used to collect the tweets to avoid any bias introduced by them being present in most of the tweets. Ten percent of the total collected data were duplicate and empty. These tweets were filtered out to obtain 866,527 unique tweets. We further tokenized the tweets using the Penn Treebank tokenizer which uses regular expressions to tokenize texts similar to the tokenization used in the Penn Treebank\footnote{https://web.archive.org/web/19970614160127/http://www.cis.upenn.edu/~treebank/}. These tokens were then passed through a Porter stemmer\footnote{https://tartarus.org/martin/PorterStemmer/} for stemming to reduce inflected word forms to their word stem for a reduction in the size of the vocabulary.


\section{Methodology}
\label{methodology}

\begin{figure}[htbp] 
\centering
\includegraphics[width=1\columnwidth]{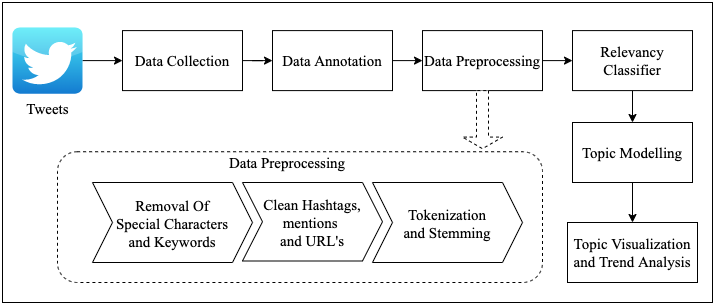}
\caption{Overview of the Framework to explore most discussed topics about the COVID-19 pandemic on Twitter.}
\label{workflow}
\end{figure}

In this section, we discuss the steps and models involved in our approach to filter the relevant tweets and understand the trending topics. Fig. \ref{workflow} presents an overview of the sequence of steps involved in our analysis. After preparing a dataset as discussed in Section \ref{sec:data_collection}, we used the 1,500 preprocessed and annotated tweets to build a relevancy classifier. Two baseline traditional machine learning algorithms using Term Frequency - Inverse Document Frequency (TF-IDF) embeddings and contextual language model embeddings were evaluated and further discussed in Sections \ref{subsec:Embedding_rep} and \ref{subsec:classification_aalgos}.  
Finally, the best performing model was selected to predict the labels of all the unique 865,027 tweets. Furthermore, we performed topic modeling on the predicted relevant tweets to estimate the probability of a tweet belonging to a unique topic and understand the hidden semantic similarities between the tweets. We discuss the details of the topic modeling algorithm further in Section \ref{subsec:Topic_algos}.

In this study, two types of classifiers were trained using two types of feature representations, and the best performing model was used as a Relevancy Classifier to infer unlabelled tweets. The performance of our supervised models was evaluated using document embeddings based on (TF-IDF) and contextual language models. A document embedding is a learned representation for a text document where words that have a similar meaning have a similar representation.

\subsection{Feature Representation} 
\label{subsec:Embedding_rep} 
Feature representation in any text classification task plays a crucial role. Traditional context-free representations such as Bag-of-Words (BoW) or TF-IDF are known for word-level representation, and in recent years we have seen the improvement in language models by including contextual semantics. We discuss various representations as follows. 

\begin{itemize}
 \item Context-Free Embedding: In a BoW based feature vector, each element in the vector represents a unique word or n-gram from the entire data set. The most significant limitation of such a representation is the inability to encode word meaning or contextual information into the vectors. TF-IDF attempts to evaluate the importance of a word by reducing the weight of words that occur more frequently in the data set and increasing the weight of words that are less frequent. The vector representation of a word based on TF-IDF, thereby encodes a different weight depending on its frequency of occurrence rather than its raw count. However, even though TF-IDF representations assign different weights to different words, they fail to capture the meaning (context) of the same word at different places (polysemy).

\item Language Models (Contextual Embedding): 
Dense distributional similarity based word embedding such as Glove, FastText, or Word2Vec bridge the gap by trying to capture the neighboring words of a particular word as its context. 
However, neighboring words cannot necessarily capture the context of a word given the complexity and variety in a sentence. Recent developments in language modeling and the release of Transformer models researchers were able to capture the context of a given word by looking at an input sequence and deciding at each step which other parts of the sequence are essential. Bidirectional Encoder Representations from Transformers (BERT) introduced a novel technique of masked language models (MLM) and next sentence prediction \cite{Devlin2018-uj} to learn the distributional contextual representation of words. 
\end{itemize}

\subsection {Relevancy Classification}
\label{subsec:classification_aalgos}
We used the 1,500 annotated tweets to build a relevancy classifier. Due to class imbalance (Relevant:77\%, Irrelevant:23\%), we used synthetic minority over-sampling technique (SMOTE) \cite{Chawla2002-st} to generate synthetic samples of minority class and train the relevancy classifier on 1,154 relevant and 346 irrelevant tweets. After oversampling, we split the dataset into train, test, and validation subsets with a distribution of  75\%, 15\%, and 10\%, respectively. 
We used a 5-fold cross-validation technique to confirm the generalizability of the model. 
Both the feature representations discussed in Section \ref{subsec:Embedding_rep} were used to train two traditional classification algorithms: Support Vector Classification (Linear SVC) and Logistic Regression (LR) to filter the relevant tweets for COVID-19 from the collected tweets. We used the Scikit-Learn\footnote{https://scikit-learn.org/stable/modules/classes.html} pipeline to generate TF-IDF vector representations. For generating contextual embeddings using pre-trained language models, the default conversion process of tokenization, and converting all sentences to a given sequence length (i.e., truncating longer sequences, and padding shorter sequences) were followed. Sentence-Transformers\footnote{ https://github.com/UKPLab/sentence-transformers} was used to generate BERT based sentence embeddings from the tweets. A pre-trained \textit{‘bert-large-uncased'} \cite{DBLP:journals/corr/abs-1810-04805} model with an embedding dimension of 768  was used to generate the sentence embeddings. We performed a Principal Component Analysis on the embeddings to reduce the dimensionality of the feature space to 300 and prevent the classifier from overfitting. 

\subsection {Topic Modeling}
\label{subsec:Topic_algos}
    Latent Dirichlet Allocation (LDA) \cite{Blei2003-qr} is an unsupervised probabilistic model to automatically identify the hidden themes (topics) in documents like tweets. It classifies the documents (i.e., tweets) into topics that are the best representation of the data set. After preprocessing the relevant tweets, as discussed in Section \ref{sec:data_preproc}, we used a coherence score to find the optimal number of topics. A coherence score provides the score of the semantic similarity between the words in a given topic \cite{Roder2015-du}. The Coherence Model\footnote{https://radimrehurek.com/gensim/models/coherencemodel.html} from Gensim\footnote{https://radimrehurek.com/gensim/} package was used for this purpose. An LDA model with the optimal number of topics as eight was inferred by the coherence score graph, as shown in Fig. \ref{coherenceplot}. The topics thus obtained consisted of the top representative words in each topic, along with their probabilities. We publish the tweet IDs with relevance labels and topics to be explored by the research community\footnote{https://github.com/preetham-salehundam/COVID-TOPIC-ANALYSIS}.


\section{Results}

\subsection{Relevancy Classifier (Research Question 1)}\label{subsec:RQ1}

The performance of Logistic Regression and Support Vector Classifier with TF IDF and BERT embeddings are reported in Table \ref{tab:hresult}. The performance of the SVC classifier with BERT embedding was better than the other settings. The best performing SVC classifier on BERT embeddings showed precision and recall of 93\% and classification accuracy of 93\% in the test set and precision and recall of 95\% with a classification accuracy of  95\% in the validation set. This model was selected to generate predictions on the entire data set of 865,027 remaining tweets (excluding the 1,500 annotated tweets). A total of 688,825 (79.6\%) tweets were classified as relevant tweets, and the remaining 176,202 were classified as irrelevant tweets. We used 689,979 relevant tweets (688,825 predicted by the relevancy classifier and 1,154 annotated as relevant earlier) for topic modeling, visualization, and trend analysis.

 



\begin{table}[htbp] 
\centering
\caption{Relevancy Classifier Results. \newline
(Acc = accuracy, F1 = F1 score, P = precision, R= recall)}
\centering 
\begin{tabular}
{|m{0.1\columnwidth}| m{0.08\columnwidth}
m{0.05\columnwidth}
m{0.05\columnwidth}
m{0.05\columnwidth} |
m{0.08\columnwidth} m{0.05\columnwidth}
m{0.05\columnwidth}
m{0.05\columnwidth}|} 
\hline 
 &\multicolumn{4}{c}{\textbf{Test Set}} 
 & \multicolumn{4}{c|}{\textbf{Validation Set}}\\  
 \hline
 &\multicolumn{8}{c|}{\textbf{TFIDF Vectors}} \\
 \hline
Model &Acc(\%)&F1(\%)&P(\%)&R(\%) &Acc(\%)&F1(\%)&P(\%)&R(\%)\\
\hline   
LR + TFIDF &90 & 90 & 91 & 90 & 93 & 93 & 92 & 93\\  
SVC + TFIDF & 87 & 87 & 88 & 87 & 89 & 89 & 89 & 88\\ 
\hline
&\multicolumn{8}{c|}{\textbf{Contextual BERT Embeddings}}\\
\hline
LR + BERT & 88 & 87 & 87& 88& 89& 89& 89& 88\\  
SVC + BERT & \textbf{93}& \textbf{92}& \textbf{93}& \textbf{92}& \textbf{95}& \textbf{95}& \textbf{95}& \textbf{95}\\
\hline  
\end{tabular}
\label{tab:hresult} 
\end{table}

\subsection{Topic Analysis (Research Question 2 and 3) } \label{subsec:RQ2_RQ3}

\begin{figure}[htbp] 
\centering

\frame{\includegraphics[width=0.8\columnwidth]{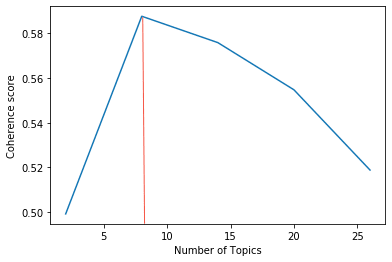}}
\caption{Coherence score graph for topic analysis with varying number of Topics. Red line shows that the coherence score was the highest when the the number of topics were 8}
\label{coherenceplot}
\end{figure}

The coherence score for varying numbers of topics for all the predicted relevant tweets is shown in Fig. \ref{coherenceplot}. As shown in the figure, the coherence score increased until eight topics and then gradually decreased; hence we chose eight to be the optimal number of topics for our data set.

The percentage of tweets discussing about each topic in the corpus is shown in Table \ref{tab:topics_distribution}. 

\begin{table}[ht] 
\caption{Topic distribution in the data set}
\centering

\begin{tabular}{| m{0.15\columnwidth} |m{0.47\columnwidth}| m{0.15\columnwidth} |} 

\hline 
 \textbf{Topic Number} &  \textbf{Theme} & \textbf{Tweets(\%)} \\ 
 \hline 
 $1$ &Pandemic impact and reopenings &11.7
 \\
 \hline
 $2$ &Government response &12.8
 \\
 \hline
 $3$ &Health workers and authorities &14.5
 \\
 \hline
 $4$ &Federal help and quarantine &6.0
 \\
 \hline
 $5$ &Origin of novel coronavirus &12.5
 \\
 \hline
 $6$ &People's thoughts during COVID-19 &21.1
 \\
 \hline
 $7$ &Case statistics &14.5
 \\
 \hline
 $8$ &Lockdown and stay at home &6.9
 \\
 
 \hline
  
\end{tabular}
\label{tab:topics_distribution} 
\end{table}
 The results of running an LDA model on all the relevant tweets for eight topics and manually curated themes summarizing the top ten most relevant terms along with some representative tweets for each topic are discussed in Table \ref{tab:topics}.

\begin{table*}[htbp]
\caption{Themes related to each topic, Words and representative tweets for each topic} 
\centering 
 \scalebox{0.8}{
\begin{tabular}{| m{0.08\columnwidth} |m{0.2\columnwidth}| m{0.4\columnwidth} | m{1.5\columnwidth} |} 

\hline 
 \textbf{Topic Number} &  \textbf{Theme} & \textbf{Words} & \textbf{Representative tweets}\\ 
\hline 

$1$ &Pandemic impact and reopenings &president, lockdown, business, american, reopen, economy, close, school, child, leader 
& 
\begin{itemize}
  \item WHO advice for international travel and trade in relation to the outbreak of pneumonia caused by a new \#coronavirus in \#Wuhan, \#China.
  \item China businesses, markets reopen, but many people stay home as coronavirus cases on rise again
\end{itemize}
\\
\hline 
$2$
&Government response
&trump, pandemic, people, governor, force, white, video, happen, america, donald
&
\begin{itemize}
  \item Unfortunately under this trump administration, I don’t believe anything they say!‘No reason for Americans to panic': White House seeks to calm fears over coronavirus.
  \item Xi Jinping has turned invisible during China's coronavirus epidemic, likely to cover his back in case things go badly wrong
  \end{itemize}
\\
\hline 
$3$
&Health workers and authorities
& state, pandemic, health, worker, response, american, youtube, public, outbreak, briefing
&
\begin{itemize}
  \item As public health officials work to respond to the Coronavirus in Washington, the number of contacts the patient came in close contact with rises to 50.
  \item World Health Organization discusses the novel coronavirus USA TODAY
  \end{itemize}
\\
\hline 
$4$
&Federal help and quarantine
&crisis, claim, give, federal, quarantine, community, high, help, life, provide
&
\begin{itemize}
  \item UK is badly prepared for recession
  \item PANIC!!!THEIR GREATEST FEAR. PUBLIC AWAKENING Bill Gates said movements towards nationalism worsened government response to the coronavirus crisis - Business Insider
  \end{itemize}
\\
\hline 
$5$
&Origin of novel coronavirus
& vaccine, house, virus, pandemic, china, april, never, social, medium, chinese
&
\begin{itemize}
  \item Appears the cause of the \#Wuhanpneumonia is as suspected, a \#novel coronavirus. Curious since it appears to have connection to seafood market if it is zoonotic in etiology.  Perfect example of why \#OneHealth and \#healthsecurity are important.
  \item Chinese government scientists have identified a new coronavirus as the cause of a mystery pneumonia outbreak in Wuhan
  \end{itemize}
\\
\hline 
$6$
&People’s thoughts during COVID-19
&people, go, think, thing, would, everyone, really, get, right, still
&
\begin{itemize}
  \item 3700 people were presumed dead from COVID-19 just like that. Something is fishy, are we just inflating figures for political reasons...
  \item We have more than enough data. We know that basically EVERY projection was wrong about the coronavirus. So....now that we have MORE THAN ENOUGH INFO...IS THE CORONA VIRUS WORSE THAN THE FLU?
  \end{itemize}
\\
\hline 
$7$
&Case statistics
& death, case, test, report, number, patient, hospital, total, update, positive
&
\begin{itemize}
  \item At least 835 coronavirus cases diagnosed in China, 25 dead
  \item \#BREAKING UK virus death toll rises by 888 to 15,464: health ministry
  \end{itemize}
\\
\hline
$8$
&Lockdown and stay at home
&realdonaldtrump, order, china, record, increase, company, stayhome, south, fight, lockdown
&
\begin{itemize}
  \item Coronavirus: China puts millions in lockdown amid rising deaths via @YouTube why Pakistan Govt does not buy screening virus equipments from China and why let people die in large.
  \item We may all have second and third waves. These can only be controlled by lockdown measures. Add the fact that Covid19 may be endemic. So if we can't find a vaccine, then we have to develop a new way of living. A new world awaits and it's not going to be good.
  \end{itemize}
\\
\hline  
\end{tabular}}
\label{tab:topics} 
\end{table*}

We discuss the themes for each topic as curated based on the top 10 relevant terms below.

\begin{itemize}

\item \textbf{Pandemic impact and reopenings:} As shown in Table \ref{tab:topics}, the keywords and representative tweets for Topic 1 represent the major impact of the COVID-19 on the economy, businesses , education, employment and other aspects of public life. It also contained tweets which discussed the possibilities and consequences of reopening certain businesses which were closed due to COVID-19 pandemic. 
\item \textbf{Government response:} Topic 2 was identified as the government response during COVID-19 pandemic which might be an indicative of the social media discussions about the response of all government agencies over COVID-19 from all over the world during the press briefings.
\item \textbf{Health workers and authorities:} During the COVID-19 pandemic,  the role of health officials, organizations and workers has been  prominent and the keywords in Topic 3 imply the same. Moreover it also covered the pharmaceutical interventions associated with COVID-19. 
\item \textbf{Federal help and quarantine:} In order to facilitate the multiple crisis scenarios caused by this global pandemic, various federal organizations came up with relief plans for proper pandemic disaster management and to help their citizens. At the same time people were quarantined to control the spread of the virus. The keywords in Topic 4 suggest the theme ‘Federal help and quarantine’ to contain such tweets. 
\item \textbf{Origin of novel coronavirus:} The keywords in Topic 5 mostly pointed towards the tweets mentioning the  origin of novel coronavirus which was believed to be the Hubei province of China, from where it spreaded all across the world.
\item \textbf{People’s thoughts during COVID-19:} Topic 6 represented a significant portion of the tweets discussing various aspects of people’s thoughts during the COVID-19 pandemic as it progressed.
\item \textbf{Case statistics:} We observed that Topic 7 could be identified as case statistics during COVID-19 as it predominantly depicted the set of tweets discussing the number of people affected by COVID-19, number of deaths and number of recovered patients etc. due to this pandemic.
\item \textbf{Lockdown and stay at home:} Lockdown and stay at home advisories had been issued during COVID-19 pandemic to reduce the exponential spread of the virus. Our topic model identified such tweets which talked about these advisories as observed by the generated keywords in Topic 8.
\end{itemize}

\begin{figure}[htbp] 
\centering

\frame{\includegraphics[width=0.8\columnwidth]{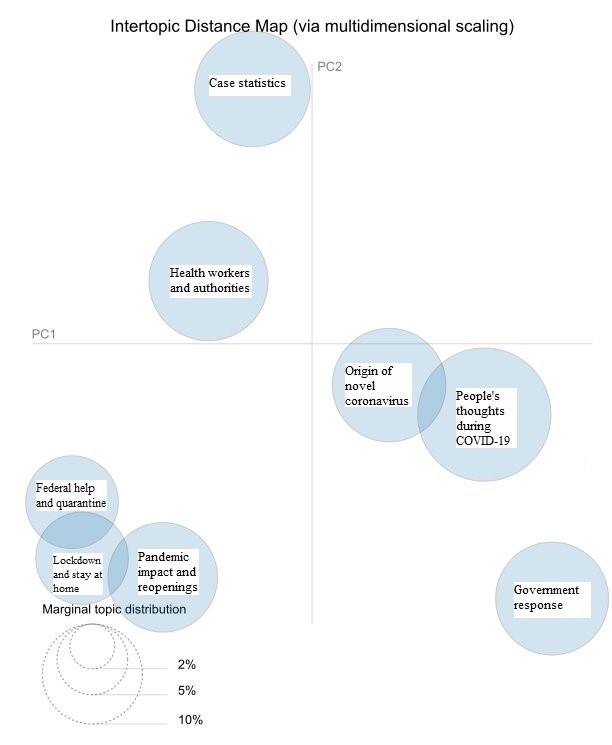}}
\caption{Topic Visualization depicting intertopic distance map. \newline PC1 and PC2 are the first two Principal Components}
\label{fig:ldaviz}
\end{figure}

An interactive visualization of the topics generated by the LDA model using pyLDAvis\footnote{https://github.com/bmabey/pyLDAvis} package as shown in Fig. \ref{fig:ldaviz} revealed that Topics 5 and 6  overlap with each other, which indicated that people's thoughts were circulating around the novel coronavirus (COVID-19) from China and theories and beliefs related to it. Topic 1, 4, and 8 also overlap with each other as these collectively address the consequences of the COVID-19 pandemic. Among all the topics, People's thoughts during COVID-19 (21.1\%), Case statistics (14.5\%), and Health workers and authorities (14.5\%) were the most discussed, which is reflected from the size of the circles in Fig. \ref{fig:ldaviz}.

\begin{figure}[htbp] 
\centering
\frame{\includegraphics[width=0.9\columnwidth]{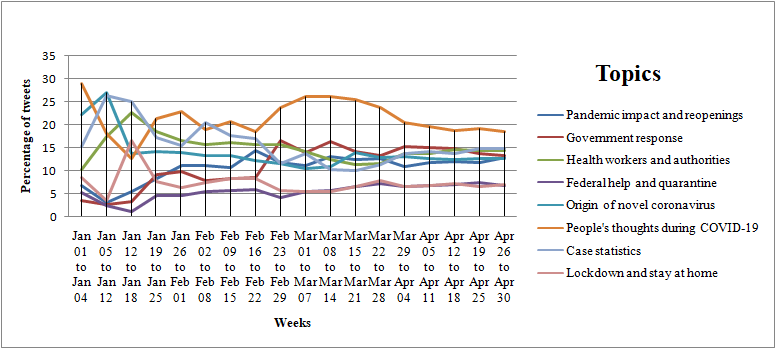}}
\caption{Topic trend with time}
\label{fig:trend}
\end{figure}

Based on the keywords in each topic, individual tweets were labeled with the prominent topic using our LDA model, and the percentage of tweets belonging to each topic for every week was calculated. As shown in Fig. \ref{fig:trend}, we plotted the percentage of tweets belonging to each theme against each week. This was done in order to visualize the trend in each topic being discussed over weeks.

\begin{figure}[htbp] 
\centering
\frame{\includegraphics[width=0.9\columnwidth]{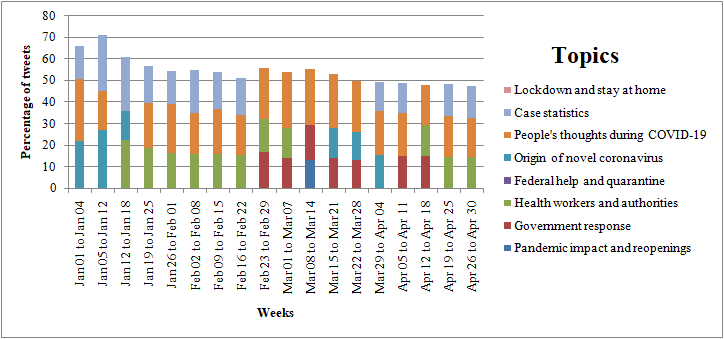}}
\caption{Prominent topics trending with time}
\label{fig:prominent_topics}
\end{figure}

In order to understand the trends in the topics reported in Fig. \ref{fig:trend}, the events occurred during COVID-19 as listed by The American Journal of Managed Care (AJMC)\footnote{https://www.ajmc.com/view/a-timeline-of-covid19-developments-in-2020} were extracted as shown in Table \ref{tab:prominent_news} (second column). Fig. \ref{fig:prominent_topics} shows the dominant topics every week, possibly due to the influence of these events. Additionally, Table \ref{tab:prominent_news} (third column) summarises the prominent topics (in terms of the percentage of tweets) each week. Such an analysis can help understand the correlation between the events and the trending topics during the COVID-19 pandemic. 

\begin{table*}[htbp]
\centering
\smallskip
\caption{Prominent event each calendar week(according to COVID-19 timeline as mentioned by AJMC) and its impact on trending topics every week}
\scalebox{0.8}{
\def\arraystretch{1.5}
\begin{tabular}{|m{3cm}| m{10cm}| m{8cm}|}
\hline
\textbf{Weeks} & \textbf{COVID-19 timeline events} & \textbf{Trending Topics} \\
\hline
 Jan 01 to Jan 11
 & WHO announced mysterious coronavirus-related pneumonia in Wuhan, China. At that time, there were 59 cases and travel precautions were already at the forefront of experts’ concerns.
 &
People’s thoughts during COVID-19, Origin of novel coronavirus and Case statistics
\\
\hline
 
Jan 12 to Jan 25
& CDC began screening for coronavirus at three US airports as cases were reported in Thailand and Japan. US also confirmed its first US coronavirus Case. Meanwhile, chinese scientist confirmed the human transmission of COVID-19 and Wuhan was placed under quarantine

&
Health workers and authorities, People's thoughts during COVID-19 and Case statistics

 \\





\hline

Jan 26 to Feb 01
& 

WHO issued a global health emergency as the worldwide death toll reached more than 200 and there was an exponential jump to more than 9800 coronavirus related cases

&  People’s thoughts during Covid-19, Health workers and authorities and Case statistics \\

\hline

Feb 02 to Feb 08
& 
Global air travel was restricted and US declared public health emergency due to the coronavirus outbreak.

&
Case Statistics, People's thoughts during COVID-19 and Health workers and authorities
\\

\hline

Feb 09 to Feb 22
& 
China’s COVID-19 deaths exceeded those of SARS crisis
& 
People’s thoughts during Covid-19, Case statistics and Health workers and authorities 
\\
\hline



 Feb 23 to Feb 29

& 
CDC said that COVID-19 was heading toward pandemic status
& 
People’s thoughts during Covid-19, Government response and Health workers and authorities

\\
\hline
 Mar 01 to Mar 07
 &
Passengers on California cruise ship were tested positive

& People’s thoughts during COVID-19, Health workers and authorities and Government response
\\
\hline
 Mar 08 to Mar 14
 & 
 WHO Declared COVID-19 a pandemic while US president Donald Trump declared it a national emergency.Travel Ban on non-US citizens traveling from Europe was enforced

& People's thoughts during COVID-19, Government response and Pandemic impact and reopenings
\\
\hline
 Mar 15 to Mar 21
 &
CMS expanded the use of telehealth temporarily and US president Trump released an economic stimulus plan to Americans. Further, California issued statewide stay-at-home order.
& People's thoughts during COVID-19, Government response and Origin of novel coronavirus
\\
\hline
 Mar 22 to Mar 28
 & Senate passed the coronavirus aid, relief, and economic security \(CARES\) Act with Trump signing it into law

& People's thoughts during COVID-19, Government response and Origin of novel coronavirus.
\\
\hline
 Mar 29 to Apr 04
 &  
 FDA authorized the use of Hydroxychloroquine for treatment of COVID-19. 

& People's thoughts during COVID-19, Government response and Origin of novel coronavirus

\\

\hline
 Apr 05 to Apr 11
 &  American Heart Association warned that the drug combination of hydroxycholorquine and antibiotic, azithromycin were not meant for everyone.
 & People's thoughts during COVID-19, Government response and Case statistics
\\
\hline
 Apr 12 to Apr 18
 & President Trump planned for the way to reopen the economy.
 
&Peoples thoughts during COVID-19, Government response and Health workers and authorities

\\
\hline
 Apr 19 to Apr 30
 & As many as 26.5 million Americans had filed unemployment by this time and due to cost concerns the poor and young people defered care for health related issues. National Institutes of Health (NIH) showed early promise for drug Remdesivir for the treatment of COVID-19
& People's thoughts during COVID-19, Case statistics and Health workers and authorities
\\ [1ex]
\hline
\end{tabular}
}
\label{tab:prominent_news}
\end{table*}

\section{Discussion}

\subsection{Principal Findings}

\subsubsection{Data Collection and Annotation Analysis}
    
    In the early weeks of January, we observed that the tweets with keywords like coronavirus and COVID-19 were very low compared to later weeks.  The volume of tweets increased from the 4\textsuperscript{th} week of January and continued to remain considerably high throughout February and decreased during April. It was also observed that the ratio of relevant to irrelevant tweets in the entire dataset (3.9) was similar to the manually labeled dataset (3.4), which shows that the distribution of the labeled data is similar to the data annotated by the classifier.
   
\subsubsection{Relevancy Classifier Analysis}
    
   It was observed that some of the tweets collected using COVID-19 specific keywords were irrelevant to the COVID-19 pandemic. We defined tweets to be relevant if they provided information about the spread, cause, effect, opinion, sentiment, emotion with regards to the COVID-19 pandemic. We designed a relevancy classifier to filter out the irrelevant tweets which was able to distinguish the tweets like “olcokcevat malnla cmertsrrnla cimri olmal veren azizsr veren zelil olur”, “nojo demais dessa pessoa”  which were tweeted using English alphabets. 
    
\subsubsection{Topic Analysis}

    Overall, among all the eight topics, People's thoughts during COVID-19, Case statistics, and Health workers and authorities were predominantly discussed. This may be due to the fact that as COVID-19 has been declared a pandemic by World Health Organization\footnote{https://covid19.who.int/}, rising cases and the role of health agencies were the main concern over social media. Several conspiracy theories were given by people \cite{Shahsavari2020-yh}, and their reactions on economic, societal, and political areas came to fore throughout the pandemic \cite{Bonaccorsi15530}. 
    In this section, we addressed the discussion on the trending topics over time by taking into consideration the COVID-19 related timeline events as listed in Table \ref{tab:prominent_news}. The key observations of the same are as discussed below:
    \begin{itemize}
        \item \textbf{Jan 01 to Jan 11:} 
        During the first two weeks of January, the World Health Organization had announced the news about the clusters of pneumonia like cases of unknown origin in Wuhan, China and travel restrictions was a concern. As a result, among all the topics discovered on our data set, "People’s thoughts during COVID-19" was the most discussed one. Tweets on Twitter during this week like "Only very limited information is available, but this looks like an outbreak of what could be a new respiratory (corona?)virus" mainly talked about some symptoms and general perceptions about this virus by the people. Topics like "Origin of novel coronavirus" and "Case statistics" were also prominent as a majority of the tweets talked about the origin of this virus and the number of people infected with it by that time.

        \item \textbf{Jan 12 to Jan 25:} By the next two weeks of January, according to the AJMC coronavirus timeline, when the cases of novel coronavirus were reported outside China in Thailand, Japan and the US, the national public health institute of US, Centre for Disease Control and Prevention (CDC) began screenings for passengers at US airports. Wuhan in China was also subsequently placed under lockdown due to reports of human to human transmission of this virus. At this time the topic, "Health workers and authorities" emerged as the dominant topics as CDC being a premier health institute started taking action to stop the spread of coronavirus. As cases of this disease started emerging in other countries outside China, the topic, "Case statistics" was also predominantly discussed at this time. 

        \item \textbf{Jan 26 to Feb 01:} During this week, as the number of COVID-19 related cases and deaths were increasing, the World Health Organization(WHO) declared a global health emergency\footnote{https://www.statnews.com/2020/01/30/who-declares-coronavirus-outbreak-a-global-health-emergency/}. So, the topic, "Health workers and authorities" continued to be discussed more as WHO is a major health organization. Tweets like "UN health agency urges China to continue search for source of new virus, as Thailand case emerges"  depicted the role of health agencies at this time. As there was also an exponential rise in the case, "Case statistics" was also one of the dominant topics.  



        \item \textbf{Feb 02 to Feb 08:}  As the US declared a health emergency due to the coroanvirus outbreak, the topic discussing about the role of "health workers and authorities" was still one of the prominent discussed topics. During this time, the topic "Case statistics" continued to be trending on Twitter at this time.

        \item \textbf{Feb 09 to Feb 22:} During these two weeks, deaths related to COVID-19 were increasing and people started talking more about this novel coronavirus. Testing and diagnosis of people for COVID-19 related symptoms by health authorities became a concern. It was
        evident from the tweet like "CDC Begins to Test Patients with Flu-like Symptoms for Coronavirus". Thus, the topics like "Case statistics" and "Health workers and authorities" were predominantly discussed at this point of time. 


        \item \textbf{Feb 23 to Feb 29:} According to CDC, as COVID-19 was heading towards a pandemic stage, "Government response" topic emerged as one of the prominent topics. Tweets like "Trump bashes media criticism over his handling of coronavirus pandemic " and "You thought the coronavirus was bad? Just look forward to the next 8 months of it being a political weapon used by the US presidential candidates" depicted the political impact of COVID-19 at this time.

        \item \textbf{Mar 01 to Mar 07:} In the first week of  March, as several passengers on a Califoria cruise ship were tested positive and world wide cases had passed 100,000, "Case statistics" topic was predominantly a topic of discussion. 

        \item \textbf{Mar 08 to Mar 14:} On March 11, WHO declared this novel coronavirus as a global pandemic. As a result, the topic, "Pandemic impact and reopenings" emerged as one of the highly discussed topics of this time. People started to discuss the effect of this pandemic as evident by the tweet "Stock market news live: Oil crashes, stock futures crater on coronavirus, crude war fears \#stocks  \#markets". Press briefings by the government officials and their response on COVID-19 became quite prominent topic on social media. 

        \item \textbf{Mar 15 to Mar 21:} By this week, US president Donald Trump signed a coronavirus relief package. 
        Several parts of the world entered into lockdown. So, "Government response" topic was highly discussed in this week. There were several conspiracy theories blaming China for such worldwide spread of the virus as evident from the tweet during this week "Is the Wuhan Coronavirus, A Bio Weapon stolen from Canada And Intentionally Released in China!". So, the discussions about the "Origin of the novel coronavirus" emerged once again as the trending topic. 

        \item \textbf{Mar 22 to Mar 28:} When the third month of the global pandemic ended, the White House and Senate leaders reached an agreement on a \$2 trillion stimulus deal. As a result "Government response" towards COVID-19 was a highly discussed topic. 

        \item \textbf{Mar 29 to April 4:} One of the United States federal executive departments, Food and drug administration (FDA) authorized the use of drug Hydroxychloroquine to treat COVID-19 on March 30. So, more discussions on "Government Response" towards COVID-19 continued on Twitter.

        \item \textbf{April 5 to April 11:} As American Health Association warned that Hydroxychloroquine along with the combination of azithromycin drugs were not meant for everyone, "People's thoughts during COVID-19" and "Case statistics" continued to trend high during this time. Meanwhile, the US president Trump began promoting this drug. Consequently, "Government response" was also trending.

        \item \textbf{April 12 to April 18:} Further, Trump started planning for the reopening of the economy according to the "gating criteria" . In addition, he announced the halt of funding to World Health Organization (WHO) as he held the organization responsible for not giving proper information about the pandemic earlier\footnote{https://www.bbc.com/news/world-us-canada-52289056}. As a result, "Health workers and authorities" topic emerged once again as the dominant topics along with "Government response". 
        
        \item \textbf{April 19 to April 30:} By the end of April, as 26.5 million people had filed unemployment in the US\footnote{https://www.cnn.com/world/live-news/coronavirus-pandemic-04-24-20-intl/h\_41202e644630af195f7adf074315e677}. According to AJMC COVID-19 timeline, people started deferring their treatments due to "cost concerns". "Health workers and authorities" was still one of the predominantly discussed topics as NIH trials had shown show the drug, remdesivir to be effective for the treatment of COVID-19. Promotion of telehealth as mentioned in the tweet "NEW: Bipartisan lawmakers back efforts to expand telehealth services for seniors to help combat the coronavirus" might have also led to the increased discussion related to "Health workers and authorities" topic during this time.


    \end{itemize}
Such analysis could be useful for policymakers, government authorities, and health officials to understand the public concerns on the events occurring during emergencies like pandemics. It could help them to provide correct and timely information on the events happening during emergencies. News reports or magazine articles need to be careful in reporting the events to the public as it might influence their discussions on social media. Otherwise, there could be widespread fake news, misinformation, and several conspiracy theories on social media platforms.

\subsection{Strengths and Limitations}

In this section, we would like to present some limitations we faced in our analysis of social media posts during the COVID-19 pandemic as below.
 \begin{itemize}

    \item \textbf{Dataset:} This study is limited to tweets in English. As the COVID-19 pandemic is a global pandemic, the English language constraint restricted our data collection to users who tweeted in the English language alone, and as a result, the findings of this study may not be generalized worldwide. We could not evaluate our analysis on multi-lingual tweets such as Hindi, German, or French, and we encourage the research community to address it in future works. Future studies may also perform spatial and temporal analysis to identify the events and topics specific to a location.

    
    \item \textbf{Geo-location:} Due to the low volume of tweets that include geotags, we have excluded it from the scope of this study. 

     \item \textbf{Fake news and misinformation:} The use of social media platforms come with a known risk of fake news and misinformation, which could affect the topics of discussion over social media. Research focused on identifying fake news would have interesting applications for understanding pandemics like COVID-19.
     \item \textbf{Event detection:} Another limitation that should be discussed is that we cannot make causal claims regarding the alignment between certain topics and events. These findings help describe the relationship between COVID-19-related timeline events and discussions on Twitter, but we cannot use these data to make causal claims. Exploration of the usefulness of Twitter data for actual detection of events related to COVID-19 is an important next step.

 \end{itemize}
\subsection{Comparison with Prior Work}
 \begin{itemize}
    \item \textbf {Relevancy Classifier:}
    During crises like natural disasters or pandemics, online social media can create noise which can clutter the useful information. In most of the prior works, the tweets collected using keywords were assumed to be relevant, and studies were carried out.  However, we found that the tweets collected using keywords can be irrelevant and contain no useful information. Twenty-five percent (175048) of the total collected tweets were irrelevant. These kinds of tweets can often bias the experiments. In order to mitigate this issue, we built a relevancy classifier that can filter out the irrelevant tweets. Most prior research has filtered out non-English tweets in the process of data preparation. However, oftentimes, we see foreign language tweets containing English alphabets included, which can create noise in the experiments. In this study, we observed that the majority of the foreign language tweets were identified as irrelevant by the relevancy classifier, and thus these were excluded from our experiments.
\item \textbf {Understanding of Events Over Time:}
    Temporal trend analysis on online social media data can be important to identify how discussions over social media get influenced over time by the news and information available to the people during pandemics like COVID-19. During events like natural disasters and pandemics, it is important for the government and humanitarian organizations to identify key factors that influence the topics of discussion of the public over time. Prior works have not addressed this problem. In this study, we identified various events that may be linked to the topics of discussion during the COVID-19 pandemic. These types of studies could also be used to understand the influence of fake news and other media sources on public discussions on social media platforms. Hence, proactive actions could be taken to stop the spread of fake news and misinformation, and at the same time, the impact of influential and impact of accurate news could be identified. 


\end{itemize}

\section {Conclusion}
While urgent actions are needed to mitigate the potentially devastating effects of COVID-19, they can be supported by understanding the behavioral and social impact on the people. Because the pandemic imposes significant psychological burdens on individuals, insights from analyzing public conversations’ trends can be used to help align public emotions with the recommendations of epidemiologists and public health experts. To understand the possible effect of media coverage on public emotions, we present an analysis of the trending topics of conversations on Twitter to events during the pandemic. Such analysis can provide insights so that a positive frame could be designed in the media coverage of the pandemic to educate the public and relieve negative emotions while increasing compliance with public health recommendations. The officials monitoring the spread of the pandemic can judge the influence of some false news or rumors related to COVID-19 as they could monitor the topics trending at a given time and thus provide appropriate information to the people. Particular sentiments like fear and misconceptions can also be monitored similarly. Government officials can make better decisions accordingly as their decisions will be reflected in the news, which would impact people’s behavior and discussions. Empirical and qualitative evaluations of our analysis indicated that our analysis is trustworthy and can inform the design of frameworks for more precise and definitive social data mining to assist humanitarian organizations during global pandemics.

\bibliographystyle{IEEEtran}
\bibliography{IEEEabrv,covid-19_references}

\end{document}